\documentclass[aps,notitlepage,twocolumn,nofootinbib,pra,superscriptaddress,10pt]{revtex4-1}
\usepackage{amsmath,amsfonts,amssymb,amsthm,bbm,graphicx,enumerate,times}
\usepackage{mathtools}
\usepackage[usenames,dvipsnames]{color}

\usepackage{hyperref}
\usepackage{tikz}
\usepackage{graphicx}
\usepackage{todonotes}
\usepackage{float}
\usepackage{comment}
	
\definecolor{jens}{rgb}{0,.8,.5}

\newcommand{\id}{\mathrm{1}}

\newcommand{\CC}{\mathbb{C}}

\newcommand{\ket}[1]{|#1\rangle}
\DeclareMathOperator{\linspan}{span}
\DeclareMathOperator{\argmin}{argmin}
\DeclareMathOperator{\decompose}{decompose}
\DeclareMathOperator{\mat}{mat}

\begin{document}
\title{Towards overcoming the entanglement barrier when simulating long-time evolution}

\author{C.\ Krumnow}
\affiliation{Dahlem Center for Complex Quantum Systems, Freie Universit{\"a}t Berlin, 14195 Berlin, Germany}

\author{J.\ Eisert}
\affiliation{Dahlem Center for Complex Quantum Systems, Freie Universit{\"a}t Berlin, 14195 Berlin, Germany}

\author{\"O.\ Legeza}
\affiliation{Strongly Correlated Systems ``Lend\"ulet'' Research Group, Wigner Research Centre for Physics, Hungarian Academy of Sciences, 1525 Budapest, Hungary}

\date{\today}

\begin{abstract}
Quantum many-body systems out of equilibrium pose some of the most intriguing questions in physics. Unfortunately, numerically keeping track of time evolution of states under Hamiltonian dynamics constitutes a severe challenge for all known methods. Prominently, tensor network methods are marred by an entanglement blowup, which allows to simulate systems following global quenches only to constant time. In this work, we present a scheme that allows to significantly extend the simulation time for interacting fermionic or equivalent spin systems. In the past when keeping track of evolution in one-dimensional real space, the subspace parametrised by real-space matrix product states satisfying area laws -- often dubbed the ``physical corner'' of Hilbert space -- was chosen as variational set. In contrast, if the manifold containing both tensor network states and fermionic mode transformations is chosen, significantly longer times can be achieved. We argue and our results suggest that in many cases it is genuine correlations between modes that is the actual limiting factor: The system at hand is for intermediate times contained in the ``physical corner'', but a different one than what is commonly assumed.
\end{abstract}
\maketitle

Some of the most intriguing problems in the study of interacting quantum many-body 
systems take their origin in the investigation of their out of equilibrium
properties. The dynamical response of strongly correlated systems to perturbations 
is dictated by such out of equilibrium properties
\cite{ngupta_Silva_Vengalattore_2011}. Conceptually 
speaking, closed quantum many-body systems following unitary 
time evolution remain equally of pivotal interest. It is known that features of
statistical mechanics are dynamically emerging in the course of such closed system
dynamics \cite{1408.5148}, bringing notions of equilibrium
statistical mechanics in contact with quantum dynamics. Still, much of
the fine-print of that mechanism and on what precise time scales 
it acts are still to be understood.
In the light of this state of affairs, it is no surprise the development of 
numerical methods that allow to study quantum many-body systems out of 
equilibrium has consistently received significant attention. 

However, all known methods are challenged by not being able to reach 
\emph{long times} in time evolution, following an initial state having short-ranged correlations (say, as is common in 
ground states of gapped phases of matter). Dynamical mean field theory
\cite{RevModPhys.86.779}, 
Monte Carlo methods \cite{RevModPhys.73.33,TroyerMonteCarlo} and -- prominently -- 
\emph{tensor network methods} \cite{Daley2004,SingleSite,Trotzky,Orus-2014,Verstraete-2008,AreaLaw} that in their range of applicability make very accurate and quantifiable predictions, all fail for long times. Indeed, this is a challenge that comes in several facets and flavours. Digging deep
into the framework of notions of computational complexity, it is also clear that 
no numerical method can reach arbitrarily long times, in that the time evolution
under translationally invariant local Hamiltonians is as powerful as a quantum
computer (and hence BQP complete in technical terms) \cite{PhysRevLett.100.010501}, 
rendering a universal efficient classical algorithm highly implausible. 
Dynamical
quantum simulators \cite{Trotzky,BlochSimulation} are for this reason also seen as
candidates that have the potential to show a quantum advantage over classical
computers.

For tensor network methods in particular -- a class of methods that provide particularly
reliable and certifiable predictions when they can be applied --
it is the \emph{entanglement entropy} growing linear in time 
\cite{SchuchQuench,AnalyticalQuench,CalabreseExtended} that 
prevents them to be applicable: 
The bond dimensions and hence variational parameters 
needed to approximate a state then grow exponentially in time
\cite{Schuch_MPS}: They leave, as one says, the ``physical corner'' of Hilbert space,
as it has been dubbed, 
the subspace that is captured by \emph{entanglement area laws} \cite{AreaLaw}. 
This means that to good precision, there is a hard
barrier that cannot be overcome in this way, even when involving supercomputers allowing for 
large bond dimensions of matrix product states \cite{Trotzky}. 

When thinking of thermalising dynamics, this gives rise to a highly ironic situation: The long-time limit can presumably be parametrized with tensor network states efficiently, in particular if one has the promise that a given many-body system indeed thermalises. From an initial low entanglement situation one is evolving the system to a situation that can again be economically captured: High temperature thermal states are again expected to be
well approximated by tensor network states \cite{Intensive,ThermalPEPOs}. 
But this process
seems to require the treatment of 
propagating excitations in a fully quantum mechanical fashion, then leading to an entanglement blowup.
Motivated by this ironic situation, truncation schemes have been suggested that lead to more economical descriptions for systems that are assumed to thermalise \cite{PhysRevB.97.035127,PollmannMPSLongTimes}. 
The downside of such an approach is that the thermalising dynamics is postulated and not found; then,
the errors made cannot be controlled.

{\it The mindset of the method.}
In this work, we suggest a novel approach to the problem at hand: We advocate the paradigm that at least part of the entanglement blowup is an artefact of representing the state vectors in a local basis. If one can adapt the basis concomitant with the time evolution, significant parts of the entanglement growth can be compensated. This is most manifest in interacting fermionic systems described by matrix product states: Here, the unitary mode transformations allow to efficiently rotate the basis in which the matrix product state is formulated, much in the spirit of established multi-configuration methods. Our scheme exploits the combination of mode transformation and matrix product states and operators on the corresponding joint manifold \cite{PhysRevLett.117.210402}: With this, when evolving the system in time, one can adapt the entanglement growth in time. This comes at the expense of the resulting Hamiltonian becoming long-ranged. Still, this leads to a verifiable
and trustworthy method that allows to accommodate significantly longer times. As a result, the genuine barrier is not the real-space entanglement, but the more slowly emerging quantum correlations between fermionic modes, a mindset expected to be particularly
applicable in regimes of small interactions, where dominating hopping terms give rise to
a quick rise of real-space entanglement entropies.

{\it Notation and definitions.}
In this work we consider systems of interaction fermions in $M$ single particle modes. To each mode $j\in[M]$ we associate the corresponding creation and annihilation operators $f^\dag_j$ and $f_j$ which fulfill the canonical anti-commutation relations
\begin{equation}
 \{f_j,f_k\} = 0,\quad\{f_j^\dag,f_k\} = \delta_{j,k}\quad\forall j,k\in[M].
\end{equation}
With $\ket{0}$ we denote the vacuum state vector containing no particles and define for $j_1,\dots,j_M\in \{0,1\}$ the \emph{Slater determinants} 
$\ket{j_1,\dots,j_M}=(f^{\dag})^{ j_1}\dots (f^\dag)^{j_M}\ket{0}$. 
We furthermore denote with $\mathcal{H}_n$ the space of all states with $n$ particles, i.e.
\begin{equation}
 \mathcal{H}_n = \linspan\left\{\ket{j_1,\dots,j_M}\ |\ \sum_k j_k = n\right\}.
\end{equation}
We denote the full Fock space as $\mathcal{F}_M = \bigoplus_{n=0}^M H_n$.
Given any unitary matrix $U\in U(M)$ we can define a new set of creation and annihilation operators according to
\begin{equation}
a_k = \sum\limits_{j=1}^M U_{k,j} f_j,\qquad a_k^\dag = \sum\limits_{j=1}^M f_j^\dag U^\dag_{j,k},
\end{equation}
which again fulfill the canonical anti-commutation relations.
Such a mode transformation corresponds to basis change in $H_1$ and acts on $\mathcal{F}_M$ via the transformation 
$\ket{\psi} \mapsto G(U)\ket{\psi}$ \cite{phd}
with $G(U)$ denoting the collection of compound matrices $G(U) = \bigoplus_{n=0}^M C_n(U)$ where $C_n(U)$ collects the determinants of all $n\times n$ sub-matrices of $U$.
Furthermore, for a state vector $\ket{\psi}\in\mathcal{F}_M$ we define the \emph{reduced one particle density matrix} (1-RDM) $\rho\in\CC^{M\times M}$ with entries
\begin{equation}
 \rho_{j,k} = \langle \psi | f_j^\dag f_k | \psi \rangle. \label{def:1RDM}
\end{equation}

A state vector $\ket{\psi}\in \mathcal{F}_M$ is called a \emph{matrix product state (MPS)}
vector with maximal bond dimension $D$ (and local dimension $d=2$), 
if it can be written as
\begin{equation}
 \ket{\psi} = \sum\limits_{i_1,\dots,i_M=0}^1 A^{[1],i_1} \dots A^{[M],i_M} \ket{i_1,\dots,i_M}
\end{equation}
with $A^{[k],i_k}\in \CC^{D_k\times D_{k+1}}$, $D_k \leq D$ for any $k\in[M]$ and $i_k\in\{0,1\}$. We consider open boundary MPS meaning that $D_1=D_{M+1}=1$
MPS feature a natural local structure as each tensor $A^{[j]}$ is associated to an individual mode. In the following, we often work on a given mode $j$. We will then assume that the MPS of interest is given in a mixed normalized form with respect to that site $j$, meaning that all tensors $A^{[k]}$ with $k<j$ and $k>j$ are left- and right-normalized \cite{Garcia2007}, 
respectively. 

For a given 4-tensor $A^{[j,j+1]}\in\CC^{2\times2\times D_j\times D_{j+2}}$ let $\sigma_1,\dots,\sigma_{2\min(D_j,D_{j+2})}$ denote the singular values of the matrix $\mat(A^{[j,j+1]})_{(i_j,\alpha),(i_{j+1},\gamma)} = A^{[j,j+1],i_j,i_{j+1}}_{\alpha,\gamma}$ then we define
\begin{equation}
 r(A^{[j,j+1]}) = \sum\limits_{k=1}^{2\min(D_j,D_{j+2})} \sigma_k = \|\sigma\|_1.\label{def:r}
\end{equation}
Furthermore, we introduce for $A^{[j,j+1]}\in\CC^{2\times2\times D_j\times D_{j+2}}$ and $U\in U(2)$ the tensor $A^{[j,j+1]}(U)$ with entries
\begin{align}
 A^{[j,j+1],i_j,i_{j+1}}(U) &=\nonumber\\ \sum_{i_j^\prime,i_{j+1}^\prime}G(U^{[j]}_{\rm  opt})&_{(i_j,i_{j+1}),(i_j^\prime,i_{j+1}^\prime)}A^{[j,j+1],i_j^\prime,i_{j+1}^\prime}.
\end{align}

{\it Mode transformations and matrix product states.}
Mode transformations and MPS have led 
to rather different mindsets when it comes to the simulation of interacting fermions -- be it the approximation of the ground state or the time evolution of a system. Mode transformations applied to a fixed Slater determinant give rise to the Hartree-Fock approximation which can be applied successfully if the system is close to be free such that the particles are the main focus. Matrix product states puts the modes in the center. A state can be efficiently written as MPS if its correlation between modes is bounded within some fixedly chosen set of modes. Methods such as the \emph{density matrix renormalisation group (DMRG)} \cite{White1992,Schollwoeck2011,Szalay-2015} or the time evolving block decimation
and variants thereof \cite{PhysRevLett.93.040502,Schollwoeck2011,Daley} allow to perform 
ground state searches and to approximate the time evolution of a given system. To both mindsets various additional methods and ramifications thereof have been developed in the past extending their applicability but the major focus of either particles or modes remains.
In Ref.~\cite{PhysRevLett.117.210402}, a method has been
developed combining mode transformations and MPS for a ground state approximation. There the single particle basis and matrices of the MPS have been optimized such that a much broader set of states is efficiently accessible to the method.
Here we show that this idea carries much further, in that a
union of both mindsets can be achieved for the time evolution of interacting fermions. The scheme presented in this work combines the time evolution method developed in Ref.~\cite{Haegeman-2014b} with the insights of Ref.~\cite{PhysRevLett.117.210402} and was developed and implemented in Ref.~\cite{phd}.

\begin{figure*}[tb]
\centering
\includegraphics[width=1\textwidth]{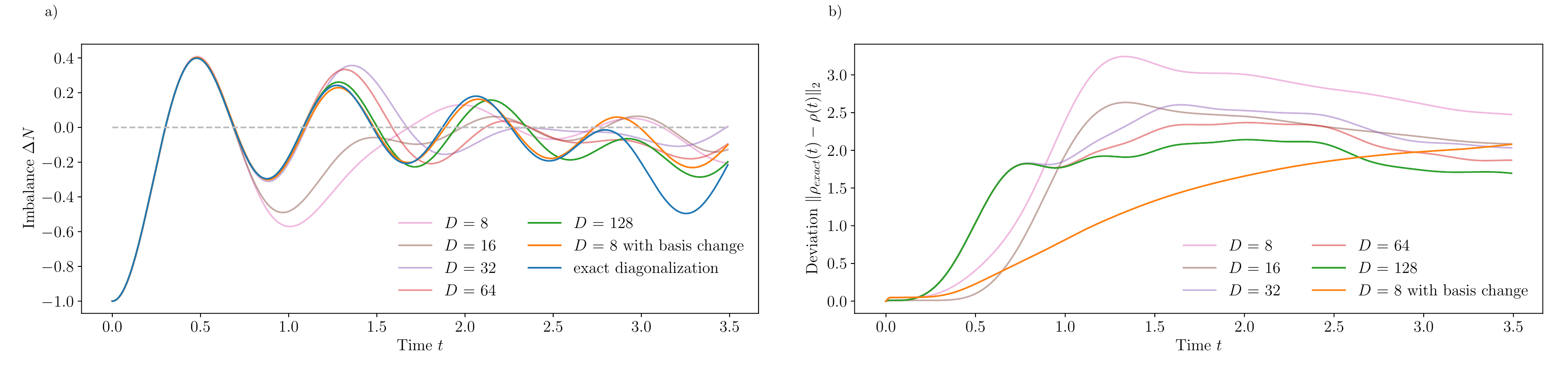}
\caption{Comparison of the time evolution in a fixed single-particle basis with increasing number of parameters to the described algorithm with a varying basis. We simulate the time evolution of the one-dimensional spin-less fermionic model defined in Eq.~(\ref{Hamiltonian}) with periodic boundary condition on $24$ sites and interaction $U = 0.25J$. As initial state vector we have chosen a charge density wave which has the occupation number representation $|\psi (0)\rangle = |1, 0, 1, 0,\dots\rangle$ in the on-site basis. We compare the exact solution in blue obtained with exact diagonalisation with the MPS calculation in the fixed real space basis with bond dimensions $D_\text{max} = 8, 16, 32,64, 128$ and a calculation in a varying basis with maximal bond dimension $D_{\text{max}} = 8$ and one mode adaption sweep after every time step. For the time evolution we used a step size $dt = 0.01$ and used a composite integrator of fourth order in all MPS calculations. In panel a), we track the imbalance of the state over time. In panel b) we show the total deviation of the 1-RDM in time from the exact solution.}
\label{Figure}
\end{figure*}

In Ref.~\cite{Haegeman-2014b} an MPS time evolution scheme for general long range Hamiltonians $H$ has been designed. One time step corresponds to one (or multiple for increased accuracy) sweeps through the MPS which gives rise to small updates of all matrices $A^{[k],i_k}(t)\mapsto A^{[k],i_k}(t+dt)$.
In order to add mode transformation let us first note that operators such as the Hamiltonian are transformed easily to the new basis as general one and two particle operators transform according to
\begin{align}
 \sum\limits_{i,j}t_{i,j} f^\dag_i f_j  &=  \sum\limits_{i,j}(U t U^\dag)_{i,j} 
 a^\dag_i a_j, \\
  \sum\limits_{i,j,k,l}v_{i,j,k,l} f^\dag_i f^\dag_j  f_k f_l &= \nonumber\\\sum\limits_{i,j,k,l}((U\otimes U)v&(U^\dag\otimes U^\dag))_{i,j,k,l} a^\dag_i a^\dag_j  a_k a_l.
\end{align}
This also applies to their pre-contracted versions used in non-local DMRG methods in order to save computation as shown in Ref.~\cite{PhysRevLett.117.210402}.

The transformation of a state vector $\ket{\psi}$ with $G(U)$ resulting from a generic $U$ 
is, however, practically infeasible in larger systems. In this context,
it is important to note that after a small time step, correlations are commonly 
created between particular modes only. Consider the example of $H$ being local on a one-dimensional chain. A small time step $dt$ will smear the correlations merely
within a support window of a size that scales with $dt$, 
resulting in only local neighbourhoods of modes becoming correlated. 
In effect, we only need to find local transformations acting on neighbouring sites in order to compensate for the built up correlations.
We do this by sweeping through the chain and optimising the modes pairwise locally. The optimization on modes $j$ and $j+1$ is performed such that the state features an as small as possible Renyi-1/2 entropy for a bi-partition into $[j]$ and $[j]^C$. 
This is a good figure of merit to ensure a more efficient MPS approximation \cite{Schuch_MPS}. 
In practice, we achieve this by minimising $r$ defined in \eqref{def:r} for pairwise blocked tensors $A^{[j]}$ and $A^{[j+1]}$. We then rotate the state with $G(U_{\rm  opt})$ -- which can be done efficiently as only the modes $j,j+1$ are mixed and therefore only the tensor $A^{[j,j+1]}$ is affected -- and counter-rotate the operators with $U_{\rm  opt}^\dag$ in order to keep all expectation values constant. The rotated tensor $A^{[j,j+1]}(U_{\rm  opt})$ is then decomposed in a way so that the MPS is in mixed normalized form with respect to the next site of the current sweep, i.e., 
either $j+1$ for a forward or $j-1$ for a backward sweep.
In order to allow for more freedom, we allow to stack multiple mode disentangling sweeps which allow to treat more general settings as well.

\begin{figure*}[tb]
\centering
\includegraphics[width=0.72\textwidth]{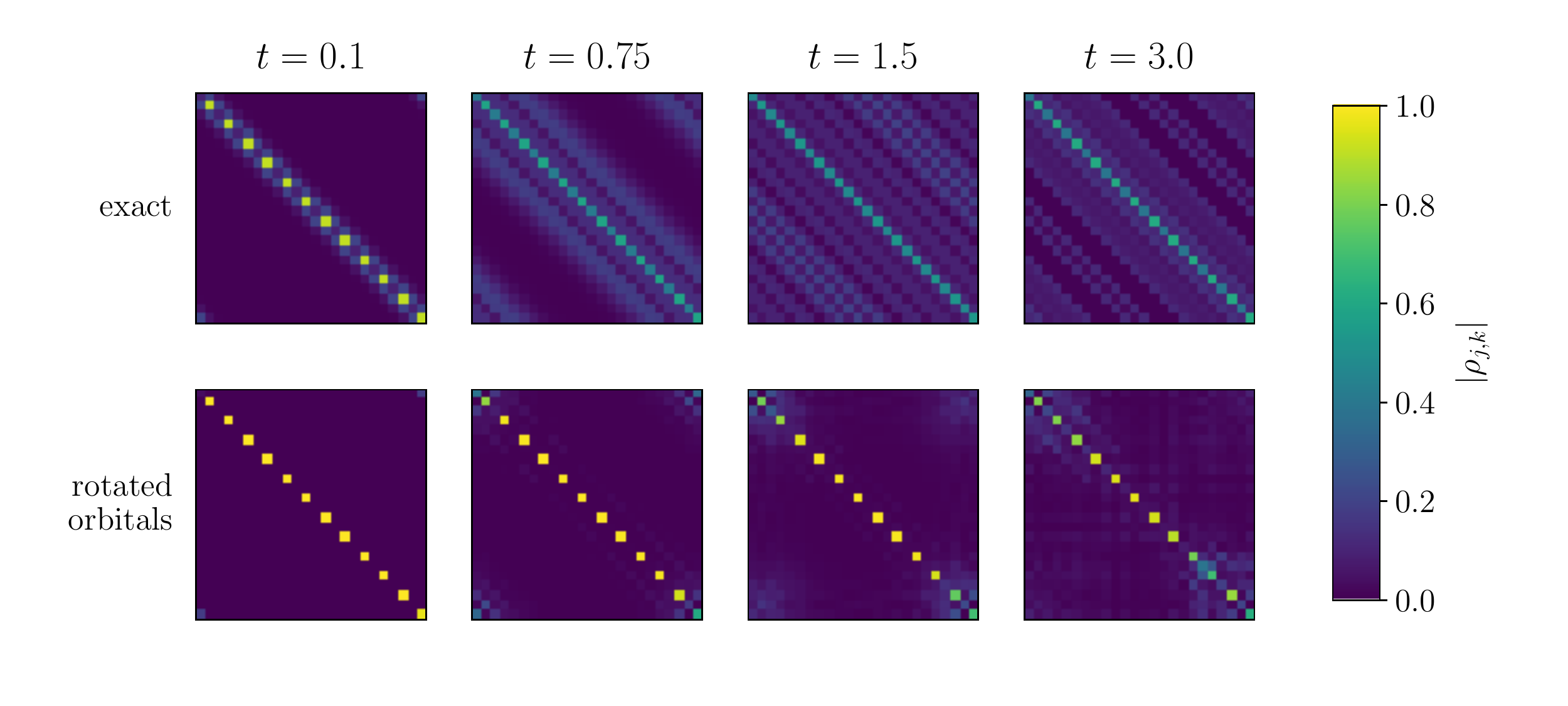}
\caption{Display of the evolution of the single particle reduced density matrix. For the computation described in Fig.~\ref{Figure}, we plot the 1-RDM at the indicated four times for the exact calculation in real space (top row) and within the rotated single particle basis (bottom row). }
\label{Figure2}
\end{figure*}

As last step, we truncate the bond dimension to the given maximum. Note that this has the consequence that when computing $A^{[j]}(t)$, we typically increase the bond dimension above $D_{\rm max}$ by a constant factor. Usually, one would directly truncate the obtained tensors back to $D_{\rm max}$, but here we allow for such a growth in order to exploit the gained residual information when adapting the single particle basis.
During the sweeps adapting the single particle basis, we do not allow the bond dimension to grow further in order to keep it bounded in the case of growing $m$. From the truncation error obtained from the decomposition and truncation of $A^{[j,j+1]}(U_{\rm  opt})$ as well as from the truncation back to $D_{\rm max}$, the total truncated weight can be computed which can be used to bound the total deviation of our approximation of $\ket{\psi(t)}$ to the true state vector. 
The resulting algorithm is summarized in Tab.\ \ref{tbl:algorithm}

\begin{table}[t]
\caption{Time evolution step with adaptive single particle basis given the Hamiltonian $H$, the initial state as an MPS $A^{[k]}(t=0)$, $t_{\rm end}$, number of mode update sweeps $m\in\mathbb{N}$ and maximum bond dimension $D_{\rm max}$ \cite{phd}.}
\label{tbl:algorithm}

{\parindent0pt\flushleft
1\quad t = 0

2\quad {\bf while} $t<t_{\rm end}$:

3\qquad $t = t+dt$

4\qquad Compute $A^{[k]}(t)$ from $A^{[k]}(t-dt)$ (as in Ref.~\cite{Haegeman-2014b})

5\qquad {\bf Iterate} $m$ times through sites $j = 1,\dots, (M-1),\dots 1$:

6\quad\qquad $A^{[j,j+1],i_j,i_{j+1}}_{\alpha,\gamma} = \sum_\beta A^{[j],i_j}_{\alpha,\beta}(t)A^{[j+1],i_{j+1}}_{\beta,\gamma}(t)$

7\quad\qquad $U_{\rm  opt} =\argmin_{U\in U(2)}(r(A^{[j,j+1]}(U)))$ (as in
Ref.~\cite{PhysRevLett.117.210402})

8\quad\qquad $A^{[j]}(t),A^{[j+1]}(t) = \decompose(A^{[j,j+1]}(U_{\rm  opt}))$ 

9\quad\qquad Transform relevant operators with $\id_{j-1}\oplus U_{\rm  opt}^\dag \oplus \id_{M-j-2}$

10\qquad Truncate bond dimension to $D_{\rm max}$

}
\end{table}

{\it Numerical example.} The method described above is very general and can be applied to any interacting fermionic Hamiltonian and is thus equally applicable to lattice models from condensed matter physics or systems treated in quantum chemistry \cite{PhysRevLett.117.210402,Szalay-2015} or nuclear physics \cite{Legeza-2015}. 
In order to illustrate its functioning and limits in most physical terms, 
we focus here on a one-dimensional real space lattice model of $L$ sites given by the Hamiltonian
\begin{equation}\label{Hamiltonian}
H = \sum_{j=1}^L \left( f_j^\dagger f_{j+1} + f_{j+1}^\dagger f_j + U 
f_j^\dagger f_j f_{j+1}^\dagger f_{j+1}\right),
\end{equation}
with the interaction strength $U$ and a hopping amplitude set to 1. It should be clear, however, that by virtue of the Jordan-Wigner transformation, one-dimensional \emph{spin models} can
be captured in the same way, treating the spin model as a model of interaction fermions. 
As initial states we use a charge density wave state vector of the form
\begin{equation}
	\ket{\psi(0)} =\ket{0,1,\dots, 0,1},
\end{equation}
resembling the situation discussed in Refs.\ \cite{Trotzky,BlochMBL} and other works. 
We then approximate the time evolution 
\begin{equation}
	\ket{\psi(t)} = e^{-itH} \ket{\psi(0)}.
\end{equation}
In reminiscence to previous numerical and experimental investigations
of relaxation processes of fermionic systems \cite{BlochMBL}, 
we investigate the evolution of the imbalance defined as
\begin{equation}
\Delta N(t) =
\frac{N_{\rm even}(t) -N_{\rm odd}(t)}{N},
\end{equation}
where $N_{\rm even}(t) = \sum_{j\ \text{even}} \langle\psi |f_j^\dag f_j |\psi\rangle$ denotes the expected particle number on all even
sites and $N_{\rm odd}(t)$ on all odd sites correspondingly.
Fig.\ \ref{Figure} shows the results for the imbalance as a function of 
time, both using standard MPS methods challenged by the entanglement blow-up
and the new method involving mode transformations. One can see
that longer times can be reached compared to
the standard MPS method: The mode transformations allows to extend the time in which the dynamics of the local observable can faithfully be reproduced significantly as the mode-correlations are reduced (compare also Fig.\ \ref{Figure2}).
Interestingly, a bond dimension as small as $D_{\rm max}=8$
is sufficient to capture the time evolution for comparably 
long times, if suitable mode transformations are allowed for. The limiting
``physical corner'' is governed by the entanglement present in the joint manifold 
including
the mode transformations, and not the real-space entanglement in the 
original basis. 
For very long times, presumably the actual obstacle of
entanglement between rotated fermionic modes sets in, defining the actual
obstacle against time evolution for infinite times.

In Fig.~\ref{Figure2} we directly compare the evolution of the 
1-RDMs. Note that for the exact calculation, $\rho$ is computed in the on-site basis $\{f^\dag_j, f_j\}$ while in case of the adapted single particle basis it is formed with the rotated operators $\{a^\dag_j, a_j\}$. 
The 1-RDM obtained from the exact computation shows a clear spreading of correlation in time.
In the rotated basis one finds that the 1-RDM stays close to diagonal,
reflecting the absence of correlations on the level of second moments of
fermionic operators, and clearly corroborating the picture of a reduction of
correlations difficult to capture by MPS by the mode transformations. Investigated in more detail, we however find, that at the start and end of the chain correlations start to build up. This is a result of the periodic boundary conditions which induce long range correlations if the sites are arranged on an open boundary chain. Such long-range contributions are harder to compensate by local transformations and are one source for the build up of correlations.

\emph{Conclusion and outlook.} In this work, we 
have challenged the picture that the growth of
real-space entanglement necessarily gives rise to the ``entanglement barrier'' that
tensor network methods following real-time evolution are unable to overcome, presumably
as the ``physical corner'' is left after an increase of certain entanglement entropies. 
Resorting to describing quantum states both by MPS and suitable mode transformations, we have identified this as possibly misleading picture, as it is not real-space entanglement that is of relevance here, but the one that limits 
a description in terms of tensor networks when expressing the state in any basis
obtained by means of a Gaussian transformation. While we have put an emphasis on fermionic
models, again, it should be clear that the same idea applies equally well to one-dimensional spin models undergoing out of equilibrium time evolution. 

Future work will provide significantly more numerical evidence for the findings laid out here. 
Equipped with these tools, questions of thermalisation and pre-thermalisation in interacting models
become freshly accessible. We expect this method to work best in the regime of
comparably small interactions, where the hopping gives rise to a large Lieb-Robinson
velocity and hence real-space entanglement buildup, in a situation in which
the mode transformations can largely accommodate for this effect.
To give further substance to the idea of modes becoming
slightly entangled in time (as opposed to real-space entanglement being relevant), future
work will also need to put two particle reduced density matrices into the focus of attention.

The work presented here offers a plethora of new research directions, once overcoming
the prejudice that real-space entanglement must be the limiting factor. 
So far the method provides an as accurate as possible descriptions of the time evolved state needed when comparing for instance theoretical and experimental predictions of non-equilibrium effects. As laid out in the introduction, such a scheme however will inevitably fail for generic interacting systems in the long time limit. The combination of the presented method and local truncation schemes might allow to formulate tensor networks describing the local thermalising dynamics efficiently. We expect this work to be relevant
both practically, describing interacting quantum systems in regimes of intermediate
times. It is also conceptually interesting, in that it provides educated advice
when aiming at delineating the boundary between systems that can be efficiently
classically described and that may show a quantum advantage of quantum 
simulators over any classical simulation method. 

\emph{Acknowledgements.} J.~E.\ has been supported by the Templeton Foundation, the ERC (TAQ), the DFG (EI 519/15-1, EI 519/14-1, CRC 183 Project B01), and the European Union’s Horizon 2020 research and innovation program under Grant Agreement
No.\ 817482 (PASQuanS). \"O.~L.\ has been supported 
by the Hungarian  National  Research,  Development  and  Innovation  Office
(NKFIH) through Grant No.\  K120569, by the Hungarian Quantum Technology National Excellence  Program (Project No.\  2017-1.2.1-NKP-2017-00001). \"O.~L.\ also  acknowledges  financial
support from the Alexander von Humboldt foundation.
Upon completion of this work, we became aware of a work that takes a different technical approach, but is close to ours mindset and motivation \cite{Rams}.


%

\end{document}